\documentclass[runningheads]{llncs}
\usepackage[T1]{fontenc}

\usepackage{graphicx}

\usepackage{rotating}

\usepackage{makecell}
\usepackage{multirow}
\usepackage{pifont}
\usepackage{xcolor}
\newcommand{\cmark}{\textcolor{green!80!black}{\ding{51}}}
\newcommand{\xmark}{\textcolor{red}{\ding{55}}}

\begin{document}
%
\title{Document Layout Annotation: Database and Benchmark in the Domain of Public Affairs}
\titlerunning{DLA: Database and Benchmark in the Domain of Public Affairs}
%

\author{Alejandro Peña\inst{1}\orcidID{0000-0001-6907-5826}, Aythami Morales\inst{1}\orcidID{0000-0002-7268-4785}, Julian Fierrez\inst{1}\orcidID{0000-0002-6343-5656}, Javier Ortega-Garcia\inst{1}\orcidID{0000-0003-0557-1948}, Marcos Grande\inst{1},
Íñigo Puente\inst{2}, Jorge Córdova\inst{2}, Gonzalo Córdova\inst{2}}

\authorrunning{A. Peña, A. Morales, J. Fierrez, et al.}
%
\institute{BiDA - Lab, Universidad Autónoma de Madrid (UAM), Madrid 28049, Spain\\
\and VINCES Consulting, Madrid 28010, Spain}
\maketitle              
\begin{abstract}
Every day, thousands of digital documents are generated with useful information for companies, public organizations, and citizens. Given the impossibility of processing them manually, the automatic processing of these documents is becoming increasingly necessary in certain sectors. However, this task remains challenging, since in most cases a text-only based parsing is not enough to fully understand the information presented through different components of varying significance. In this regard, Document Layout Analysis (DLA) has been an interesting research field for many years, which aims to detect and classify the basic components of a document. In this work, we used a procedure to semi-automatically annotate digital documents with different layout labels, including $4$ basic layout blocks and $4$ text categories. We apply this procedure to collect a novel database for DLA in the public affairs domain, using a set of $24$ data sources from the Spanish Administration. The database comprises $37.9$K documents with more than $441$K document pages, and more than $8$M labels associated to $8$ layout block units. The results of our experiments validate the proposed text labeling procedure with accuracy up to $99\%$.

\keywords{Document Layout Analysis  \and Legal domain \and Data Curation \and Natural Language Processing .}
\end{abstract}
\section{Introduction}

Nowadays, the Portable Document Format (PDF), originally developed by Adobe and standardized~\cite{pdf_iso} in $2008$, has become one of the most important file formats for digital document storing and sharing. The reason behind this success is the possibility to present documents including a variety of components (e.g. text, multimedia content, hyperlinks, etc.) in a format independent from the software, hardware and operating system. Furthermore, this file format allows encryption, compression, digital signature, and even interactive editing (e.g. form filling).

The advantages of the PDF format have converted it in a basic document tool for governments, administrations or enterprises. However, despite its usefulness, automatic processing of digital PDF documents remains as a difficult task. To correctly process and extract information from a document, it is required first to understand how the different components of the document are structured and how they interact with each other. For instance, processing information contained in a table usually requires to previously detect its basic structure. Even when it comes to text processing, text blocks in documents can be grouped into a variety of semantic levels (e.g. body text, titles, captions, etc.), which have different relevance and presentation formats. The way in which basic elements are presented in a document to effectively transmit its message is known as document layout. Once the document layout is clear, then modern Natural Language Processing (NLP) technologies (e.g., transformers~\cite{kenton2019bert}\cite{brown2020gpt3} with attention mechanisms~\cite{vaswani2017attention}) can be applied for generating useful outputs from segmented text blocks. 

\begin{figure}[t!]
    \centering
    \includegraphics[width = \textwidth]{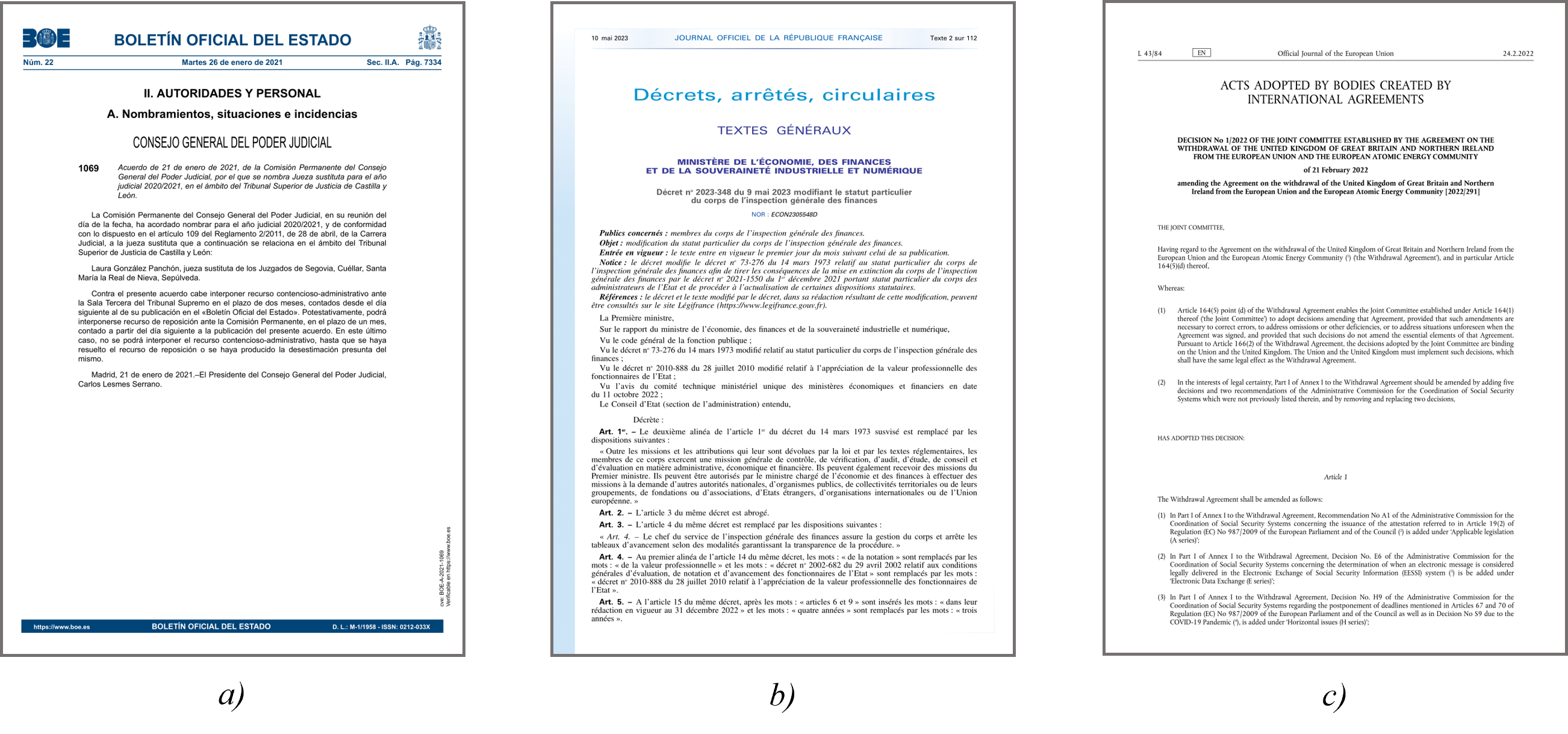}
    \caption{Visual examples of page document images from different official gazettes: a) spanish gazette (i.e., BOE); b) french gazette (i.e., Légifrance); and c) Official Journal of the European Union.}
    \label{fig:layout_examples}
\end{figure}

Document Layout Analysis (DLA) is a task that aims to detect and classify the basic components of a document. As we previously introduced, this task is a crucial component within the automatic document processing pipeline. Nevertheless, its usefulness is proportional to its difficulty. The main reason behind this fact is the large variability inherent in the problem. In this work, we propose a method to semi-automatically annotate a large number of digital PDF documents with their basic layout components. Our method combines a document collection procedure, the use of PDF miners to extract layout information, as well as a human-assisted process of data curation. We use this pipeline to generate a corpus of official documents for DLA in the legislative domain, which we call Public Affairs Layout (PAL) database. The source of the documents in this work are official gazettes from different institutions of the Spanish Administration. Official gazettes are periodical publications,\footnote{https://op.europa.eu/en/web/forum/european-union} in which administrations include legislative/judicial information and announcements. Despite the fact that they originate from different administrations and countries, these documents usually present common features related with the spatial location and visual characteristics of the different text blocks. Take for example the document page images presented in Figure~\ref{fig:layout_examples}. While Figure~\ref{fig:layout_examples}.a is an example of a spanish official gazette, its layout it's similar to that of Figure~\ref{fig:layout_examples}.b and Figure~\ref{fig:layout_examples}.c, where page images from the french and the EU official gazettes are respectively depicted. Independently of the language of the document, a reader can easily identify the different text blocks (e.g., titles, body, summary). 

In this context, the main contributions of this work are the following:
\begin{itemize}
    \item We present Public Affairs Layout (PAL) database,\footnote{https://github.com/BiDAlab/PALdb} a new publicly available dataset for DLA, collected from a set of $24$ different legislative sources from official organisms. The database comprises nearly $37.9$K documents, $441$K pages, and more than $8$M layout labels.
    \item We provide layout information extracted from the documents, including the pre-processed cleaned text from the text blocks detected. Thus, in addition to the DLA dataset, a large corpus of public affairs text in spanish, and other 4 co-official languages, was collected for NLP pre-training and domain adaptation.
    \item We assess our text labeling strategy with different experiments, in which we prove the usefulness of the information extracted to classify text blocks into different semantic classes defined after an empirical analysis of the data sources.

\end{itemize}

The rest of the paper is structured as follows: Section~\ref{sec:related_work} provides a review of different works concerning document layout analysis and automatic digital document processing. In Section~\ref{sec:dataset} we present our semiautomatic procedure to collect and annotate our legislative document database, as well as its details. Section~\ref{sec:experiments} presents the experiments and results of this work. Finally, Section~\ref{sec:conclusions} summarizes the main conclusions.

\section{Related Work}
\label{sec:related_work}
The literature on Document Layout Analysis (DLA) differences between two types of PDF documents: $1$) native or digital-born documents and $2$) document image. The former are those originally created from a digital version of the documents, while the latter are scanned images captured from a physical document, or digital-born documents which were converted to images. This distinction leads to different approaches on how to extract their main layout components.

With regard to PDF native documents, the availability of all document information within the internal PDF structure makes the use of PDF miner tools the most common approach to extract layout information. A large number of tools exist for layout extraction in digital documents. However, for long time these tools were mainly focused on text extraction~\cite{pdf_benchmark}, and were limited by the way the PDF format processed their components (it specifies where and how to place individual components, without using high level semantic information about them). This behavior also makes difficult to detect layout elements such as tables, as there is no label to identify them. Modern PDF miners had learn to work with this structure, but elements such as tables remain difficult to detect. We can cite here the work of Bast \textit{et al.}~\cite{pdf_benchmark}, where an evaluation of $14$ text extraction tools, including their own, was conducted. They also proposed a benchmark for text extraction methods from digital PDF documents, and collected a database consisting on $12$K scientific articles from \textit{arXiv}, which they annotated by parsing the corresponding TeX files. More recently, the authors of~\cite{zhong2019publaynet} proposed an automatic method to annotate a large corpus of digital PDF articles, by matching the output of the \textit{PDFMiner\footnote{https://github.com/euske/pdfminer}} library with the XML representations of the articles. They released a page image database, known as PubLayNet, and later made available the original native PDF documents used to create it.

On the other hand, DLA on document images has been addressed as an image processing task with the use of Computer Vision techniques. By not having access to internal information of the documents, especially when working with scanned documents, databases in this domain were mostly annotated at hand, which ultimately limited their size. For instance, we can cite here the datasets collected for the ICDAR document processing challenges~\cite{antonacopoulos2009icdar}\cite{clausner2015icdar}\cite{clausner2017icdar}, which included complex documents with realistic layouts, the ICDAR $2013$ Table Recognition Challenge dataset~\cite{gobel2013icdar}, or the UW III and UNLV datasets. Other works present their own manually annotated databases based on scientific articles~\cite{oliveira2017fast}\cite{soto2019visual}, with the previously mentioned PubLayNet~\cite{zhong2019publaynet} being the larger database (i.e. nearly $350$K page images) thanks to an automatic annotation method.

Early approaches for conducting DLA on document images included text segmentation techniques~\cite{bukhari2011improved}\cite{eskenazi2017comprehensive} or the use of HoG features~\cite{lang2018physical}\cite{sah2017text} to perform the task. More recently, deep learning-based methods have been applied, specially the use of R-CNN object detection models~\cite{he2017mask}\cite{ren2015faster}. In~\cite{soto2019visual} a combination of F-RCNN~\cite{ren2015faster} with contextual information was proposed to perform this task. The authors of PubLayNet~\cite{zhong2019publaynet} used both F-RCNN and M-RCNN~\cite{he2017mask} in their experiments on the novel database. Oliveria \textit{et al.} proposed~\cite{oliveira2017fast} the use of $1$-D CNNs as both an efficient and fast solution for DLA. They used the running length algorithm~\cite{wong1982document} to detect regions of information in grayscale images, and detect blocks as regions connected after a $3\times3$ dilatation operation. Then, the network classifies the blocks using both vertical and horizontal projections of these blocks.

\section{Semiautomatic Document Layout Annotation}
\label{sec:dataset}

In this Section, we will present the Public Affairs Layout (PAL) Database, a new database for DLA in the legal domain, with special focus in official documents. More specifically, Section~\ref{sec:dataset_collection} presents the data sources, our data collection method, and the details of the final database. Section~\ref{sec:layout_extraction} describes the tools used to extract layout information from the documents, the features extracted, and the different layout components of the database. Finally, in Section~\ref{sec:data_curation} we explain our semiautomatic method to classify the text blocks extracted into different semantic categories, and the following data curation procedure.

\subsection{Document Collection}
\label{sec:dataset_collection}

Spain has a wide variety of sources of public affairs documents. All the judicial, royal and governmental decrees, as well as the laws approved by the congress, have to be published in the daily official gazette. Our data collection involves $24$ public sources including $3$ national administrations, $19$ regional administrations, and $2$ local administrations in charge of the different territories in which Spain is politically/administrative divided (i.e. autonomous communities/cities). There are 5 co-official languages in the spanish territory, and each region has the freedom to publish official documents in their own format. All of this generates a great variety of styles and formats.  

We included in our database a total $24$ different sources of information from official organisms. The main characteristics of these data sources are summarized as follows:

\begin{enumerate}
    \item The availability of historical repositories of PDF files with more than ten years of almost daily publications (i.e. there are usually no publications on Sundays).
    \item The diversity of document layouts, which is different for each source and it has been changing over the years for each one.    
\end{enumerate}

Table~\ref{tab:data_sources} in the Appendix presents the full list of data sources used in this work. Since we had access to historical repositories of all the publications, we used an automatic web scrapping method to download all the documents. We then filtered documents published before $2014$, and used the most current ones in our work. This allowed us to discard scanned-image PDF files corresponding to old publications, which were left out of the scope of this work. We use as web scraping backbone the Python library Scrapy\footnote{https://scrapy.org/}, concretely the Spider class, where we defined how the website would be parsed.

\begin{table}[t]
\resizebox{\textwidth}{!}{
    \centering
    \begin{tabular}{l|c|c|c|c|c|c|c|c|c|c}
    \hline
    \multirow{2}{*}{\textbf{Source ID}}&\multirow{2}{*}{\textbf{\#Doc.}}&\multirow{2}{*}{\textbf{\#Pages}}&\multirow{2}{*}{\textbf{\#Tokens}}&\multicolumn{7}{c}{\textbf{Layout Components}}\\
    \cline{5-11}
    &&&&\textbf{\#Images}&\textbf{\#Tables}&\textbf{\#Links}&\textbf{\#ID}&\textbf{\#Title}&\textbf{\#Summary}&\textbf{\#Body}\\
    \hline\hline
    $1$&$193$&$602$&$182.5$K&$1206$&$179$&$1$&$2296$&$2463$&$192$&$5540$\\
    \hline
    $2$&$16$&$231$&$143.3$K&$0$&$0$&$0$&$715$&$1040$&$11$&$2138$\\
    \hline
    $3$&$14$&$224$&$48.9$K&$19$&$150$&$0$&$743$&$1369$&$62$&$2562$\\
    \hline
    $4$&$28$&$857$&$329.5$K&$80$&$141$&$0$&$2735$&$6287$&$199$&$9518$\\
    \hline
    $5$&$50$&$403$&$176.6$K&$845$&$106$&$2$&$810$&$1647$&$114$&$5293$\\
    \hline
    $6$&$30$&$884$&$189$K&$65$&$449$&$105$&$3034$&$3108$&$144$&$6718$\\
    \hline
    $7$&$122$&$393$&$205.9$K&$2$&$93$&$1$&$786$&$1695$&$128$&$7166$\\
    \hline
    $8$&$44$&$649$&$299.1$K&$5593$&$176$&$3$&$793$&$1702$&$283$&$7052$\\
    \hline
    $9$&$96$&$570$&$139.1$K&$6$&$346$&$103$&$1709$&$1189$&$102$&$5114$\\
    \hline
    $10$&$13$&$1046$&$736.8$K&$1023$&$279$&$268$&$275$&$5394$&$214$&$13.3$K\\
    \hline
    $11$&$75$&$476$&$170$K&$102$&$141$&$3$&$928$&$1496$&$76$&$4277$\\
    \hline
    $2$&$41$&$742$&$367.3$K&$725$&$128$&$145$&$2232$&$5146$&$251$&$15.7$K\\
    \hline
    $13$&$43$&$310$&$118$K&$600$&$17$&$27$&$930$&$727$&$55$&$3742$\\
    \hline
    $14$&$43$&$281$&$131.2$K&$865$&$98$&$275$&$774$&$1164$&$42$&$3670$\\
    \hline
    $15$&$50$&$225$&$69.5$K&$8$&$38$&$0$&$659$&$852$&$50$&$2737$\\
    \hline
    $16$&$13$&$383$&$204.4$K&$1064$&$140$&$22$&$382$&$1209$&$73$&$4100$\\
    \hline
    $17$&$142$&$315$&$143.9$K&$7$&$72$&$372$&$941$&$1283$&$157$&$4899$\\
    \hline
    $18$&$35$&$1064$&$297.2$K&$12304$&$273$&$5$&$2127$&$1410$&$224$&$9766$\\
    \hline
    $19$&$10$&$1302$&$532.5$K&$1511$&$4384$&$0$&$2592$&$1962$&$470$&$15.4$K\\
    \hline
    $20$&$40$&$1049$&$348.3$K&$1049$&$63$&$0$&$2098$&$1951$&$44$&$32.1$K\\
    \hline
    $21$&$32$&$887$&$323.8$K&$862$&$114$&$268$&$1779$&$2320$&$293$&$15.8$K\\
    \hline
    $22$&$57$&$549$&$453.3$K&$626$&$309$&$0$&$547$&$1941$&$210$&$4534$\\
    \hline
    $23$&$61$&$4771$&$1.45$M&$15468$&$1919$&$413$&$9608$&$12563$&$882$&$48.9$K\\
    \hline
    $24$&$197$&$1064$&$454.8$K&$1071$&$17$&$23$&$1049$&$2006$&$282$&$14.4$K\\
    \hline\hline
    \textbf{Total}&$\mathbf{1444}$&$\mathbf{19276}$&$\mathbf{7.52M}$&$\mathbf{45.1K}$&$\mathbf{9632}$&$\mathbf{2036}$&$\mathbf{40.5K}$&$\mathbf{61.9K}$&$\mathbf{4558}$&$\mathbf{244.4K}$\\
    \hline
    \end{tabular}}
    \caption{Description of the PAL database (validation set). We provide statistics on the number of documents, pages and tokens for each source, along with the number of examples of each layout category.}
    \label{tab:data_val}
\end{table}

The PAL database comprises $37$,$910$ documents, in which we have $441.3$K document pages and $138.1$M tokens. Attending to the layout labels, we can find $1$M images, $118.7$K tables, $14.4$K links, and $7.1$M text blocks. The database is divided into a train set, and a validation set, where text labels were validated by a human supervisor, as we will comment later in Section~\ref{sec:data_curation}. The train set is composed of $36$,$466$ documents, $422$K document pages and $130.5$M tokens, with $1.1$M images, $145.2$K tables, $16.3$K links and $8.8$ M text blocks. The list of sources we used to collect our data, as well as the number of documents, pages, tokens and examples from each layout category included in our validation set are presented in Table~\ref{tab:data_val}. We set a minimum of $10$ documents and $200$ pages for each data source in the validation set. Note that some sources present significant differences in the pages per documents relation (e.g. for Source $1$ the relation is roughly $3$ pages per document, while for Source $19$ is $130$). This is due to the nature of the documents we had access to. While all the documents we downloaded are official gazettes, these gazettes were available in two different formats: $1$) full gazette contained in one document, and $2$) the gazette divided in different, individual documents, each one containing a section or announcement. Most gazettes are available in both formats, but the access we had to these using the Spiders varies between publications. Another interesting relation is the number of tokens vs the number of pages, which is significantly higher for two sources, namely Source $22$ (i.e. $825.7$ tokens per page) and Source $10$ (i.e. $704.4$ tokens per page). These publications are the only ones among the $24$ to present a two-columns format, so the number of tokens per page is naturally higher.

All the documents we collected are PDF native format, that is, they are not scanned images of a document, rather they were originally created from a digital version of the document. This fact allowed us to use PDF miner tools to identify the main document layout components, and extract information related to these elements. In the following Section, we will introduce the different layout components extracted in this work, as well as our semiautomatic annotation algorithm.

\subsection{Layout Components Extraction}
\label{sec:layout_extraction}

We considered $4$ main layout categories or \textit{blocks} in this work: $1$) image; $2$) table; $3$) link (i.e. a region in the document associated to an external URL), and $4$) text blocks. We used $2$ different PDF miner libraries to extract these components from the documents:

\begin{itemize}
    \item \textbf{PyMuPDF}.\footnote{https://pymupdf.readthedocs.io/en/latest/} This is a Python binding for MuPDF, a powerful PDF viewer and toolkit. We use this tool to extract image, link, and text blocks from each page of the documents. PyMUPDF not only allows us to detect these blocks (i.e. returning their position as a $4$-tuple bounding box ($x_0$, $y_0$, $x_1$, $y_1$)), but also returns information about them (e.g. the raw text of the blocks, font type, size, etc.)
    \item \textbf{Camelot}.\footnote{https://pypi.org/project/camelot-py/} A Python library to extract tables from PDFs. This library detects the position of the tables in each page by getting both the vertical and horizontal lines composing the table, and computing their boundaries (again as a $4$-tuple bounding box). Then, it extracts their information in a \textit{pandas dataframe} that preserves their structure, which can be exported in different formats.
\end{itemize}

It's important to note that these libraries work only for PDF native documents, therefore any document image scanned included in them was treated as a simple image. For each input PDF file, we extract the layout information and the information from all the tables detected. We also generate a version of the input PDF file annotated with layout information, which allowed us to visually assess the results of the extraction.

\begin{figure}[htbp]
    \centering
    \includegraphics[width = 0.7\textwidth]{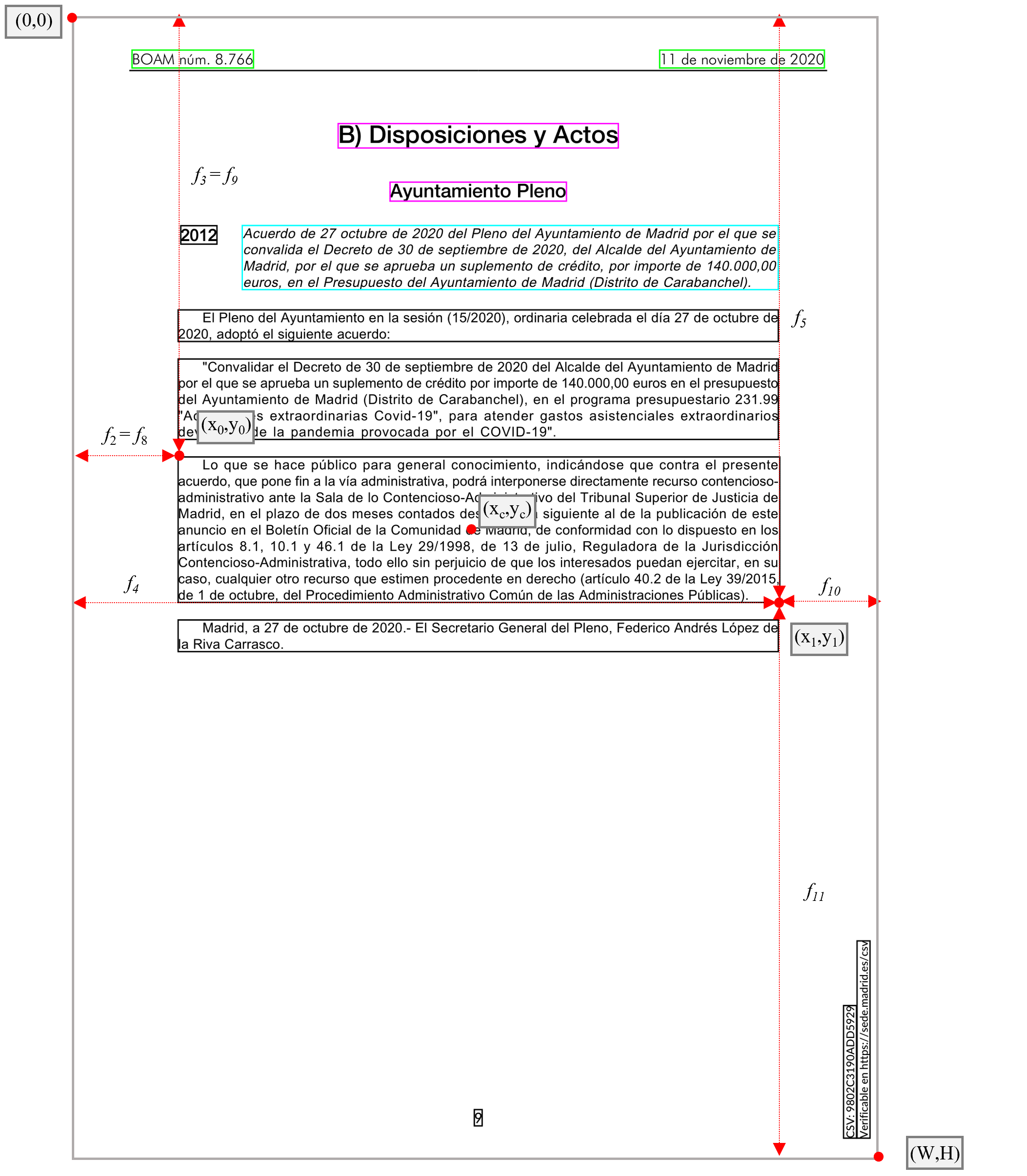}
    \caption{Document page image with Layout component annotations (\textit{L} in Table~\ref{tab:features}). The color codification is green for \textit{Identifier}, pink for \textit{Title}, cyan for \textit{Summary}, and black for \textit{Body}. We also illustrate the coordinate system and the different positional features for an example text block.}
    \label{fig:coordinate_system}
\end{figure}

When extracting tables with Camelot, we adapted the bounding boxes to PyMuPDF's coordinate system, which considers the origin ($0$, $0$) in the top left corner (see Figure~\ref{fig:coordinate_system}). We also had to consider the case of rotated tables (i.e. wide tables that appear rotated to fit in a whole page), which presented another coordinate system different from normal tables.

\begin{table*}[htbp]
\resizebox{\textwidth}{!}{
    \centering
    \begin{tabular}{l c|c|c|c|c}
    \hline
    \multirow{2}{*}{\textbf{Feature}}&\multirow{2}{*}{\textbf{Description}}&\multicolumn{4}{c}{\textbf{Layout Blocks}}\\
    \cline{3-6}
    &&\textbf{Image}&\textbf{Table}&\textbf{Link}&\textbf{Text}\\
    \hline\hline
    $f_1$&Page number in which the block was detected, starting from $0$&\cmark&\cmark&\cmark&\cmark \\
    \cline{3-6}
    $f_{2-5}$&\makecell{A $4$-tuple ($x_0$, $y_0$, $x_1$, $y_1$) bounding box that defines the block's region \\in the page (see the coordinate system in Figure~\ref{fig:coordinate_system})}&\cmark&\cmark&\cmark&\cmark \\
    \cline{3-6}
    $f_{6-7}$&A $2$-tuple ($x_c$, $y_c$), where $x_c = (x_0 + x_1)/2$, $y_c = (y_0 + y_1)/2$&\cmark&\cmark&\cmark&\cmark \\
    \cline{3-6}
    $f_{8-11}$&\makecell{A $4$-tuple with the block distance to each \\limit of the page (see the coordinate system in Figure~\ref{fig:coordinate_system})}&\cmark&\cmark&\cmark&\cmark \\
    \cline{3-6}
    $f_{12}$&\makecell{Important data about the block (output CSV file path for tables,\\ URL for links, and the pre-processed text for text blocks)}&\xmark&\cmark&\cmark&\cmark \\
    \cline{3-6}
    $f_{13}$&Proportion of bold tokens in the text block&\xmark&\xmark&\xmark&\cmark \\
    \cline{3-6}
    $f_{14}$&Proportion of italic tokens in the text block&\xmark&\xmark&\xmark&\cmark \\
    \cline{3-6}
    $f_{15}$&Average font size of the text block&\xmark&\xmark&\xmark&\cmark \\
    \cline{3-6}
    $f_{16}$&A tuple with the different font types in the text block&\xmark&\xmark&\xmark&\cmark \\
    \cline{3-6}
    $f_{17}$&Proportion of capital letters in the text block&\xmark&\xmark&\xmark&\cmark \\
    \cline{3-6}
    $f_{18}$&Number of elements separated by simple space in the text block&\xmark&\xmark&\xmark&\cmark \\
    \cline{3-6}
    $B$&Type of block detected ($0$ for image, $1$ for table, $2$ for link, and $3$ for text)&\cmark&\cmark&\cmark&\cmark \\
    \cline{3-6}
    $L$&\makecell{Layout component label ($0$ for image, $1$ for table, $2$ for link, $3$ for identifier, \\$4$ for title, $5$ for summary, and $6$ for main text)}&\cmark&\cmark&\cmark&\cmark \\
    \hline
    \end{tabular}}
    \caption{Layout features extracted for each layout block detected by our algorithm.}
    \label{tab:features}
\end{table*}

We extracted the features presented in Table~\ref{tab:features} for the $4$ layout blocks studied in this work. As we commented before, thanks to PyMUPDF's tools we had access to different font characteristics from the text blocks, including size, font type, bold or italic information. This allowed us to define several features describing the text blocks. Note here that our approach is limited to the data contained in the PDF structure of the document. Hence, the feature extraction depends on this information, and ultimately on the editor used to create the files. We extracted text blocks in reading order (i.e. following a top-left to bottom-right schema) at line level, merging close lines with similar font features (except for size, which we averaged using the number of tokens of each size in the resulting blocks). In this step, we pre-processed the raw text to remove line breaks, excessive white spaces, and $\backslash$uFFFF Unicode characters, which were found in substitution of white spaces in the raw text extracted from Source $5$ documents. It's worth to mention here the case of Source $4$ documents, where the raw text appeared without white spaces when trying to extract it. This was probably due to the original PDF editor, which instead of using white spaces just put each word in its corresponding place. We could extract each word individually with PyMuPDF, and reconstruct the original text using their block and line references. Finally, we removed any text blocks with an overlap over $70\%$ with a table detected.

Note that not all the text blocks have the same semantic role within a document. Some document components, such as tables and images, usually have a clearly defined purpose. However, a text block may be a paragraph inside the body text, a title, or a caption, among other options. Furthermore, these semantic roles of the text are usually denoted in a document by using different layout features (e.g. the use of bold or italics, variety of fonts, different sizes, specific positions, etc.). For this reason, we inspected the documents from each source, and defined $4$ different text categories within them:

\begin{enumerate}
    \item \textbf{Identifier}. A text block that identifies the document. Here we can find the date, the number of the publication, or even source-specific identifiers.
    \item \textbf{Summary}. A text block that can be found at the beginning of some announcements, and summarizes their content.
    \item \textbf{Title}. A text block which identifies different sections within the document, or has a significant higher importance than the body text. They usually present different font characteristics than regular text blocks.
    \item \textbf{Body}. The text blocks composing the main content of the document.
\end{enumerate}

Some visual examples of these text categories can be found in Figure~\ref{fig:coordinate_system}. Considering the text categories, layout components in our documents are labeled into $1$ among $7$ possible categories. As these text categories have a semantic meaning in our documents, but they might not have it for some other applications, we made an explicit distinction between \textit{Block} (B) and \textit{Label} (L) annotations in our dataset (see Table~\ref{tab:features}). Note that for tables, links and images, the value of both annotations are the same. 

\subsection{Text Block Labeling and Data Curation}\label{sec:data_curation}

\begin{figure}
    \centering
    \includegraphics[width = \textwidth]{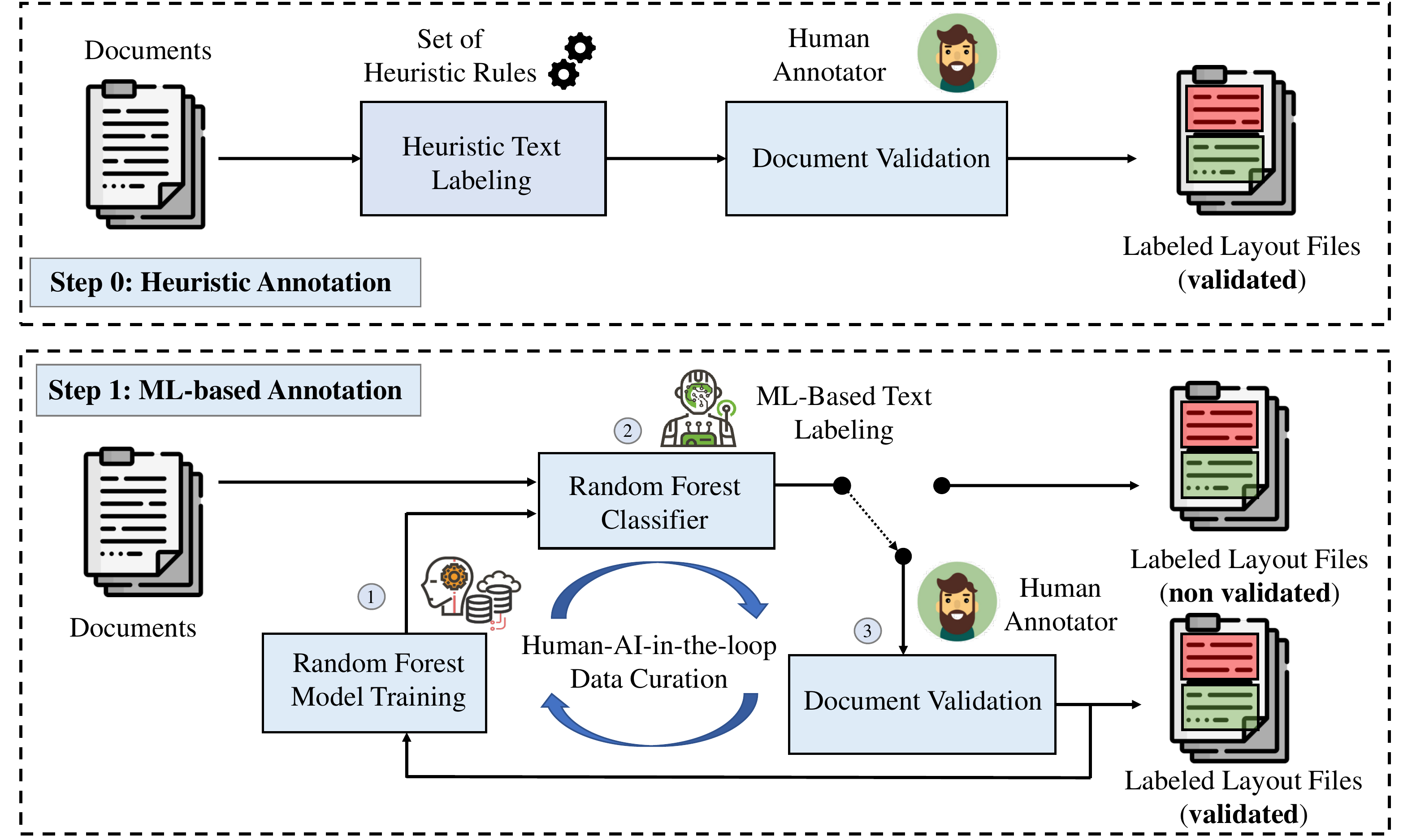}
    \caption{Text block label's data curation process. During Step $0$ we assign labels to text blocks using a set of heuristic rules, which are validated and corrected by a human supervisor. Then, we use the validated documents to train a text labeling classifier per source, hence reducing the number of errors and accelerating the labeling process.}
    \label{fig:data_curation}
\end{figure}

As we introduced in the previous Section, we defined $4$ different text categories for the text blocks of our documents. After extracting the layout components into layout files, we had a set of $18$ different features describing each text block. We proceeded to annotate each text block based on the features in a two-step process, which is illustrated in Figure~\ref{fig:data_curation}.

In the first step, or initialization step (Step $0$ in Figure~\ref{fig:data_curation}), we defined a set of heuristic rules after an initial inspection of the feature values for each source and text category (e.g. text blocks with a proportion of bold tokens over $0.5$ are labeled as \textit{Title}). The goal of the heuristic rules wasn't to perform a perfect labeling of the text blocks, but to have an initial set of documents with noisy labels. We defined these rules based only in the font text features ($f_{13}$-$f_{18}$ in Table~\ref{tab:features}), except the font types ($f_{16}$) and the presence of some key words in the text ($f_{12}$) to detect the identifiers (e.g. the Spanish words for the days of the week). We developed an application to help a human supervisor validate the resulting noisy-labeled documents, and correct wrong labels. For an input file, this application displayed each page annotated with the current bounding boxes and labels. By clicking inside a bounding box the supervisor switch to the block label to the next value. This allowed us to obtain an initial set of correctly labeled documents for each document source in a ``quick'' way. All the documents were validated by a unique human supervisor, with the aim of preventing subjective biases of different supervisors from creating a disparity in the labeling strategy (specially when it came to title labeling). However, the resulting validated documents were assessed by the different authors, who agreed with the labeling criteria.

Once we had the initial set of validated documents for a specific source, we moved to the next step (Step $1$ in Figure~\ref{fig:data_curation}). We considered a threshold of $50$ pages as the minimum set to proceed. In this step, we started by training a classifier for each source with their validated documents. We decided to use a Random Forest classifier for this task, as the nature of our rule-based decision making labeling was close to the hierarchical logic behind such classifier. 

We qualitatively assessed that, by using the trained models to validate new documents, the number of errors significantly dropped. Our experiments in Section~\ref{sec:experiment_classification} will demonstrate quantitative results supporting this fact. Hence, from this point we continue with a Human-in-the-loop AI-based Data Curation process. The use of the AI-based labeling speeded up the validation process significantly, as many errors made by the rule-based labeling were corrected. This allowed us to increase the validated sets, and retrain the models with more data, which ultimately ended up reducing further the errors. We repeated periodically this iterative process, and end up with models whose outputs were satisfactory. At this point, we use the AI-labeling process  to create a large set of unvalidated data whose labels were clean enough for a training set. A visual inspection of an arbitrary selection of documents assessed this hypothesis.

\section{Experiments}
\label{sec:experiments}
In this Section, we present the main experimental setup and results of this work. Our experiments aim to validate the usefulness of the layout features extracted from the text blocks to train text labeling classifiers.

\subsection{Automatic Text Blocks Labeling}
\label{sec:experiment_classification}

As we previously introduced in Section~\ref{sec:data_curation}, we applied a data curation method based on the use of a text labeling classifier to annotate the text blocks in our database. In this Section, we report an experiment to quantitative assess our strategy, and the usefulness of the text features extracted (see Table~\ref{tab:features}) to discriminate between different semantic text categories. Recalling Section~\ref{sec:layout_extraction}, we defined $4$ different text categories in our work after an initial inspection of the data sources: $1$) \textit{Identifier}; $2$) \textit{Title}; $3$) \textit{Summary}; and $4$) \textit{Body Text}. Among these, the most common one is the \textit{Body Text} category, with \textit{Summary} being the least frequent (see Table~\ref{tab:data_val}).

During our text labeling procedure (presented in previous Section \ref{sec:data_curation}), an individual classifier was trained for each document source using the validated documents. Here, we will evaluate the performance of these individual classifiers. We chose a Random Forest (RF) model as classifier with a maximum depth of $1000$, inspired by the low number of features and the hierarchical decision logic of the problem, for which we consider the RF model to be the perfect suit. We use as input for the classifiers all the features depicted in Table~\ref{tab:features}, except for feature $f_{12}$ (i.e. the raw text data itself). We normalized features $f_2$ - $f_{11}$ with respect to the dimensions of the document. For feature $f_{16}$, we created a dictionary to assign a value to each unique configuration of text fonts (remember that we can find different font types within the same text block, which we stored as a tuple). 

\begin{table}[t!]
    \centering
    \begin{tabular}{l|c|c|c|c|c}
    \hline
    \multirow{2}{*}{Source ID}&\multicolumn{5}{c}{\textbf{Accuracy (\%)}}\\
    \cline{2-6}
    &Overall&ID&Title&Summary&Body\\
    \hline\hline
    $1$&$98.08_{1.17}$&$99.54_{0.41}$&$96.20_{4.35}$&$99.47_{1.05}$&$98.48_{1.18}$\\
    \hline
    $2$&$96.59_{3.61}$&$100.0_{0.00}$&$91.57_{6.10}$&$100_{0}$&$97.99_{2.36}$\\
    \hline
    $3$&$98.26_{1.66}$&$100_{0}$&$96.72_{3.53}$&$97.75_{4.53}$&$98.91_{0.08}$\\
    \hline
    $4$&$99.03_{1.00}$&$99.99_{0.02}$&$98.33_{1.81}$&$96.55_{4.68}$&$99.24_{0.78}$\\
    \hline
    $5$&$96.26_{2.67}$&$100_{0}$&$94.03_{4.95}$&$100_{0}$&$96.22_{3.73}$\\
    \hline
    $6$&$95.93_{2.61}$&$98.67_{3.51}$&$88.48_{7.07}$&$90.98_{6.44}$&$98.92_{0.80}$\\
    \hline
    $7$&$97.90_{1.02}$&$100_{0}$&$96.56_{1.76}$&$99.20_{1.60}$&$97.87_{1.61}$\\
    \hline
    $8$&$94.81_{2.03}$&$100_{0}$&$86.39_{9.85}$&$89.08_{4.67}$&$97.62_{1.55}$\\
    \hline
    $9$&$97.19_{1.61}$&$100_{0}$&$89.10_{6.56}$&$98.52_{2.26}$&$98.66_{1.48}$\\
    \hline
    $10$&$98.70_{0.95}$&$99.85_{0.30}$&$97.60_{2.47}$&$96.84_{4.48}$&$99.26_{0.46}$\\
    \hline
    $11$&$96.54_{2.89}$&$99.96_{0.13}$&$96.02_{4.06}$&$98.75_{2.50}$&$95.45_{4.21}$\\
    \hline
    $12$&$99.35_{0.55}$&$100_{0}$&$98.03_{1.95}$&$98.16_{2.66}$&$99.74_{0.42}$\\
    \hline
    $13$&$98.25_{1.91}$&$100_{0}$&$93.90_{6.84}$&$97.98_{4.07}$&$98.74_{1.52}$\\
    \hline
    $14$&$96.52_{1.95}$&$99.69_{0.63}$&$94.65_{5.36}$&$100_{0}$&$96.27_{3.17}$\\
    \hline
    $15$&$94.64_{3.27}$&$99.33_{1.18}$&$89.13_{7.40}$&$93.09_{6.37}$&$95.54_{3.56}$\\
    \hline
    $16$&$93.37_{3.20}$&$99.89_{0.34}$&$84.35_{10.72}$&$61.08_{12.35}$&$96.67_{2.30}$\\
    \hline
    $17$&$97.16_{0.83}$&$100_{0}$&$93.25_{2.99}$&$97.37_{2.00}$&$97.68_{0.84}$\\
    \hline
    $18$&$98.94_{0.91}$&$99.93_{0.09}$&$94.95_{4.23}$&$97.96_{3.49}$&$99.35_{0.79}$\\
    \hline
    $19$&$98.00_{1.21}$&$99.91_{0.19}$&$90.50_{3.53}$&$99.58_{0.65}$&$98.89_{1.09}$\\
    \hline
    $20$&$99.93_{0.07}$&$100_{0}$&$99.44_{1.47}$&$99.09_{2.73}$&$99.95_{0.05}$\\
    \hline
    $21$&$99.38_{0.52}$&$100_{0}$&$96.61_{3.14}$&$93.96_{5.88}$&$99.76_{0.45}$\\
    \hline
    $22$&$98.22_{1.28}$&$98.87_{3.38}$&$96.87_{2.39}$&$95.03_{5.55}$&$98.90_{0.84}$\\
    \hline
    $23$&$97.36_{1.01}$&$99.96_{0.04}$&$91.08_{3.65}$&$99.55_{0.47}$&$98.25_{1.62}$\\
    \hline
    $24$&$99.27_{0.44}$&$100_{0}$&$95.25_{3.97}$&$99.63_{0.75}$&$99.79_{0.06}$\\
    \hline
    \end{tabular}
    \caption{Overall and per class accuracies of the text block classifiers for each data source. We report the accuracy in terms of $mean_{std}$ over a $10$ Folds Cross Validation protocol.}
    \label{tab:experiment_1}
\end{table}

We use $80\%$ of the documents from the validation set (see Table~\ref{tab:data_val}) for training, and $20\%$ for testing. Note that we decided to make the train/test splitting at document level, instead of page level, so we could take into account potential intra-document biases (i.e. the existence of significant layout differences within a document with respect to the classic source's template, where annexes, for instance, play an important role). For each source, we repeated the experiment $10$ times with arbitrary train/test splits, and report as result the mean and the standard deviation of both overall and per-class accuracies. 

The results are presented in Table~\ref{tab:experiment_1}. We report both the overall accuracy and the accuracy per class. We can observe that the best results per class are those obtained for \textit{Identifier}, with a mean accuracy over $99\%$ in all cases except for two sources, Source $6$ and Source $22$. These two cases also show a standard deviation higher than $3$, which further highlights the increased difficulty in detecting identifiers in these sources. The good results obtained for the \textit{Identifier} class could be expected, as these text blocks present low variability in the documents (i.e. they usually have the same position in the documents, number of tokens, and font characteristics). After them, the \textit{Body} class presents the best results, with all the mean accuracies over $95\%$ and low standard deviations in general. On the other hand, the ``worst'' performance is obtained for the \textit{Title} category. We consider this class as the less objective during labeling, as different considerations of what is a title can be correct, especially when encountered within the body text of the announces, or in the annexes. This subjective nature makes titles harder to classify. Another reason behind this performance is the potential mistakes between \textit{Title} and \textit{Body text} classes, as the difference in the presence of these classes penalizes errors more for the former. Nevertheless, the mean performance in the \textit{Title} class is over $84\%$ in all cases, with $19$ sources reaching a performance over $90\%$. Attending to the overall accuracies, all the sources obtain good results, which surpass $96\%$ for $20$ different sources. The lowest performance is obtained in the documents from Source $16$, which also shows an outlier-like performance in the summaries (i.e. $61.08\%$) and the lowest performance in titles (i.e. $84.35\%$).

\section{Conclusions}
\label{sec:conclusions}

In this work, we developed a new procedure to semi-automatically extract layout information from a set of digital documents, and provide annotations about the main layout components. Our methods are based on the use of web scraping tools to collect documents from different pre-defined sources, and extract layout information with classic PDF miners. The miners not only detect tables, links, images, and text blocks in the documents, but provide us with different information about these blocks, including font characteristics of the text blocks. We then defined a set of text features, which are useful to describe the text blocks and discriminate between semantic categories.

We applied our procedure to generate a new DLA database composed of public affairs documents from spanish administrations sources. After an initial inspection of the documents from each source, we defined $4$ different text categories, and classify the text blocks in these categories using a Human-in-the-loop AI-based Data Curation process. Our data curation process trains text labeling models using human-validated documents in an iterative way (i.e. the output of the models are validated and corrected by a human supervisor, leading to new validated documents to train more accurate models).

Our experiments assessed the usefulness of the text features to discriminate between the previously defined classes, thus validating our data curation procedure. We then explored the use of Random Forests to train a classifier per source, whose results validated the proposed strategy. As future work, we suggest exploring other text labeling models, such as recurrent models, which could  allow the flow of information between blocks in the page. Other text semantic classes could be studied, depending on the nature of the source documents, or potential applications after the DLA such as topic classification \cite{penna2023icdar}. Finally, the scanned images left out in this work could be processed using Computer Vision models trained with the PAL database, and added in future versions of the database. Multimodal methods for integrating native digital and image-based information for improved DLA will be also investigated \cite{2023_SNCS_multiAI}, in addition to analysis \cite{deAlcala2023n-sigma} and compensation \cite{Serna2022ai} of possible biases in the machine learning processes involved in our developments.

\section{Acknowledgments}

Support by VINCES Consulting under the project VINCESAI-ARGOS and \linebreak BBforTAI (PID2021-127641OB-I00 MICINN/FEDER). The work of A. Peña is supported by a FPU Fellowship (FPU21/00535) by the Spanish MIU.

%
%
%
\bibliographystyle{splncs04}
\bibliography{bibliography}

\begin{thebibliography}{10}
\providecommand{\url}[1]{\texttt{#1}}
\providecommand{\urlprefix}{URL }
\providecommand{\doi}[1]{https://doi.org/#1}

\bibitem{antonacopoulos2009icdar}
Antonacopoulos, A., et~al.: A realistic dataset for performance evaluation of
  document layout analysis. In: ICDAR. pp. 296--300 (2009)

\bibitem{pdf_benchmark}
Bast, H., Korzen, C.: {A benchmark and evaluation for text extraction from
  PDF}. In: 2017 ACM/IEEE Joint Conference on Digital Libraries (2017)

\bibitem{brown2020gpt3}
Brown, T., et~al.: Language models are few-shot learners. Advances in Neural
  Information Processing Systems  \textbf{33},  1877--1901 (2020)

\bibitem{bukhari2011improved}
Bukhari, S., et~al.: Improved document image segmentation algorithm using
  multiresolution morphology. In: Document Recognition and Retrieval XVIII.
  vol.~7874, pp. 109--116 (2011)

\bibitem{clausner2015icdar}
Clausner, C., et~al.: {The ENP image and ground truth dataset of historical
  newspapers}. In: ICDAR. pp. 931--935 (2015)

\bibitem{clausner2017icdar}
Clausner, C., et~al.: {ICDAR2017 competition on recognition of documents with
  complex layouts-RDCL2017}. In: ICDAR. vol.~1, pp. 1404--1410 (2017)

\bibitem{deAlcala2023n-sigma}
DeAlcala, D., Serna, I., Morales, A., Fierrez, J., et~al.: Measuring bias in
  {AI} models: An statistical approach introducing {N-Sigma}. In: COMPSAC
  (2023)

\bibitem{eskenazi2017comprehensive}
Eskenazi, S., et~al.: A comprehensive survey of mostly textual document
  segmentation algorithms since 2008. Pattern Recognition  \textbf{64},  1--14
  (2017)

\bibitem{gobel2013icdar}
G{\"o}bel, M., et~al.: {ICDAR 2013 table competition}. In: Proceedings of the
  International Conference on Document Analysis and Recognition. pp. 1449--1453
  (2013)

\bibitem{he2017mask}
He, K., et~al.: {Mask R-CNN}. In: Proceedings of the IEEE International
  Conference on Computer Vision and Pattern Recognition. pp. 2961--2969 (2017)

\bibitem{pdf_iso}
{Document management — Portable document format — Part 1: PDF 1.7}.
  Standard, International Organization for Standardization (ISO) (July 2008)

\bibitem{kenton2019bert}
Kenton, J., et~al.: {BERT: Pre-training of deep bidirectional transformers for
  language understanding}. In: Proceedings of NAACL-HLT. pp. 4171--4186 (2019)

\bibitem{lang2018physical}
Lang, T., et~al.: Physical layout analysis of partly annotated newspaper
  images. In: Proceedings of the 23rd Computer Vision Winter Workshop. pp.
  63--70 (2018)

\bibitem{oliveira2017fast}
Oliveira, D., Viana, M.: {Fast CNN-based document layout analysis}. In:
  Proceedings of the IEEE International Conference on Computer Vision. pp.
  1173--1180 (2017)

\bibitem{penna2023icdar}
Peña, A., Morales, A., Fierrez, J., et~al.: Leveraging large language models
  for topic classification in the domain of public affairs. In: ICDAR (2023)

\bibitem{2023_SNCS_multiAI}
Peña, A., Serna, I., et~al.: Human-centric multimodal machine learning: Recent
  advances and testbed on {AI}-based recruitment. SN Computer Science
  \textbf{4} (2023)

\bibitem{ren2015faster}
Ren, S., et~al.: {Faster R-CNN: Towards real-time object detection with region
  proposal networks}. Advances in Neural Information Processing Systems
  \textbf{28} (2015)

\bibitem{sah2017text}
Sah, A., et~al.: {Text and non-text recognition using modified HOG descriptor}.
  In: Proceedings of the IEEE Calcutta Conference. pp. 64--68 (2017)

\bibitem{Serna2022ai}
Serna, I., et~al.: {Sensitive Loss}: Improving accuracy and fairness of face
  representations with discrimination-aware deep learning. Artificial
  Intelligence  \textbf{305} (2022)

\bibitem{soto2019visual}
Soto, C., Yoo, S.: Visual detection with context for document layout analysis.
  In: EMNLP-IJCNLP. pp. 3464--3470 (2019)

\bibitem{vaswani2017attention}
Vaswani, A., et~al.: Attention is all you need. Advances in Neural Information
  Processing Systems  \textbf{30} (2017)

\bibitem{wong1982document}
Wong, K., et~al.: Document analysis system. IBM Journal of Research and
  Development  \textbf{26}(6),  647--656 (1982)

\bibitem{zhong2019publaynet}
Zhong, X., et~al.: {PubLayNet: Largest dataset ever for document layout
  analysis}. In: ICDAR. pp. 1015--1022 (2019)

\end{thebibliography}

\section*{Annex}
\renewcommand{\thetable}{A\arabic{table}}

\setcounter{table}{0}

\begin{table}[]
\resizebox{\textwidth}{!}{
    \centering
    \begin{tabular}{l|c|c|c}
    \hline
    \textbf{Data Source}&\textbf{ID}&\textbf{Language}&\textbf{Access}\\
    \hline\hline
        Boletín Oficial del Estado&$1$&Spanish&boe.es/diario\_boe\\
         \hline
        Boletín del Congreso de los Diputados&$2$&Spanish&congreso.es/indice-de-publicaciones\\
         \hline
        Boletín del Senado&$3$&Spanish&\makecell{senado.es/web/actividadparlamentaria/\\publicacionesoficiales/senado/boletinesoficiales}\\
         \hline
         \hline
        Boletín de la Comunidad de Madrid&$4$&Spanish&bocm.es\\
         \hline
        Boletín de la Rioja&$5$&Spanish&web.larioja.org/bor-portada\\
         \hline
        Boletín de la Región de Murcia&$6$&Spanish&borm.es\\
         \hline
        Boletín del Principado de Asturias&$7$&Spanish&sede.asturias.es/ast/servicios-del-bopa\\
         \hline
        Boletín de Cantabria&$8$&Spanish&boc.cantabria.es/boces/\\
        \hline
        Boletín Oficial del País Vasco&$9$&Spanish, Basque&euskadi.eus/y22-bopv/es/bopv2/datos/Ultimo.shtml\\
        \hline
        Boletín de Navarra&$10$&Spanish, Basque&bon.navarra.es/es\\
         \hline
        Boletín de la Junta de Andalucía&$11$&Spanish&juntadeandalucia.es/eboja.html\\
         \hline
         Boletín de Aragón&$12$&Spanish&boa.aragon.es\\
         \hline
        Boletín de Islas Canarias&$13$&Spanish&gobiernodecanarias.org/boc\\
         \hline
        Boletín de Islas Baleares&$14$&Spanish, Catalan&caib.es/eboibfront/\\
         \hline
        Boletín de Castilla y León&$15$&Spanish&bocyl.jcyl.es\\
         \hline
        Boletín de la Ciudad de Ceuta&$16$&Spanish&ceuta.es/ceuta/bocce\\
         \hline
        Boletín de Melilla&$17$&Spanish&bomemelilla.es/bomes/2022\\
         \hline
         Diario de Extremadura&$18$&Spanish&doe.juntaex.es/\\
        \hline
        Diario de Castilla$-$La Mancha&$19$&Spanish&docm.jccm.es/docm/\\
         \hline
         Diario de Galicia&$20$&Galician&xunta.gal/diario-oficial-galicia/\\
         \hline
        Diari de la Generalitat Valenciana&$21$&Spanish, Valencian&dogv.gva.es/es\\
         \hline
        Diari de la Generalitat Catalana&$22$&Spanish, Catalan&dogc.gencat.cat/es/inici/\\
         \hline
         \hline
        Boletín del Ayuntamiento de Madrid&$23$&Spanish&\makecell{sede.madrid.es/portal/site/tramites/\\menuitem.944fd80592a1301b7ce0ccf4a8a409a0}\\
         \hline
        Boletín del Ayuntamiento de Barcelona&$24$&Spanish, Catalan&\makecell{w123.bcn.cat/APPS/egaseta/\\home.do?reqCode=init}\\
         \hline
         
    \end{tabular}}
    \caption{List of data sources used to collect the PAL Database.}
    \label{tab:data_sources}
\end{table}

\end{document}